\documentclass[review,3p]{elsarticle}

\makeatletter
\def\ps@pprintTitle{%
   \let\@oddhead\@empty
   \let\@evenhead\@empty
   \let\@oddfoot\@empty
   \let\@evenfoot\@empty
}
\makeatother

\usepackage{lineno}
\modulolinenumbers[5]
\usepackage[dvipsnames]{xcolor}
\usepackage{amsmath}
\usepackage{pdfpages}
\usepackage[para]{footmisc}
\usepackage{soul}
\usepackage{xcolor}
\definecolor{lightyellow}{cmyk}{0,0,0.50,0}
\sethlcolor{lightyellow}

\usepackage{subcaption}
\usepackage{multirow}
\usepackage{textcomp}
\usepackage{booktabs}
\usepackage{comment}
\usepackage{enumitem}
\usepackage{adjustbox}
\usepackage{threeparttable}
\usepackage{array}
\usepackage{booktabs}
\usepackage{tabularx}
\usepackage[hidelinks]{hyperref}
\usepackage[nameinlink, capitalise, noabbrev]{cleveref}

\date{September 2022}

\begin{document}

\begin{frontmatter}

\title{Coordinated Inauthentic Behavior and Information Spreading\\on Twitter}

\author[sap]{Matteo Cinelli\fnref{emailmc}}
\author[iit]{Stefano Cresci\corref{mycorrespondingauthor}}
\cortext[mycorrespondingauthor]{Corresponding author}
\ead{stefano.cresci@iit.cnr.it}
\author[sap]{Walter Quattrociocchi\fnref{emailwq}}
\author[iit]{Maurizio Tesconi\fnref{emailmt}}
\author[iit]{Paola Zola\fnref{emailpz}}
\address[iit]{Institute of Informatics and Telematics, National Research Council (IIT-CNR), Italy}
\address[sap]{Dept. of Computer Science, Sapienza University of Rome, Italy}
\fntext[emailmc]{matteo.cinelli@uniroma1.it}
\fntext[emailwq]{quattrociocchi@di.uniroma1.it}
\fntext[emailmt]{maurizio.tesconi@iit.cnr.it}
\fntext[emailpz]{paolazola90@gmail.com}

\begin{abstract}
We explore the effects of coordinated users (i.e., users characterized by an unexpected, suspicious, or exceptional similarity) in information spreading on Twitter by quantifying the efficacy of their tactics in deceiving feed algorithms to maximize information outreach. 
In particular, we investigate the behavior of coordinated accounts within a large set of retweet-based information cascades identifying key differences between coordinated and non-coordinated accounts in terms of position within the cascade, action delay and outreach. On average, coordinated accounts occupy higher positions of the information cascade (i.e., closer to the root), spread messages faster and involve a slightly higher number of users. 
When considering cascade metrics such as size, number of edges and height, we observe clear differences among information cascades that are associated to a systematically larger proportion of coordinated accounts, as confirmed by comparisons with statistical null models.
To further characterize the activity of coordinated accounts we introduce two new measures capturing their infectivity within the information cascade (i.e., their ability to involve other users) and their interaction with non-coordinated accounts. Finally, we find that the interaction pattern between the two classes of users follows a saturation-like process. A larger-scale targeting of non-coordinated users does not require a larger amount of coordinated accounts after a threshold value $\sim50\%$, after which involving more coordinated accounts within a cascade yields a null marginal effect. Our results contribute to shed light on the role of coordinated accounts and their effect on information diffusion.
\end{abstract}

\end{frontmatter}


\section{Introduction}
\label{sec:introduction}
Social media play a pivotal role in the evolution of public debates and in shaping public opinion. Platforms that were originally designed for entertainment are nowadays one of the main gateways to information sources and the principal environment in which public opinions get shaped. Users can access and share an unprecedented amount of information, and feed algorithms affect the selection process. Such a constant information overload determines a new tendency of news consumption \citep{bakshy2012role,flaxman2016filter,schmidt2017anatomy}, and generates opportunities for civic engagement and public interest \citep{hagen2020rise}, also eliciting the so-called ``democratizing effect''. However, social media are also affected by a number of ailments, including toxicity, harassment and hateful speech~\mbox{\citep{trujillo2022make,zinovyeva2020antisocial}}, polarization and radicalization~\mbox{\citep{santos2021link}}, propaganda and misleading information~\mbox{\citep{hristakieva2022spread}}. Regarding the latter issue, many studies recently investigated the spread of fake information and propaganda~\citep{cinelli2020limited,yuan2021improving} and the role of disinformation on social media~\citep{del2016spreading,lazer2018science,zhang2020overview}. Within this context, scholars also investigated the role of automated accounts in spreading both reliable and questionable information~\citep{shao2018spread,mendoza2020bots}. This literature highlights that whether initial social bots could be accurately identified by state-of-the-art algorithms, recent manipulation campaigns carried out with modern bots and with human-operated accounts (i.e., trolls) are more sophisticated and thus harder to detect~\citep{boneh2019relevant,starbird2019disinformation}.

Going beyond early literature that focused on the \textit{nature} (automated or otherwise) of the accounts as a way to distinguish between malicious and legitimate actions, more recent research moved the focus on the \textit{actions} themselves and on the \textit{behaviors} of the accounts. In a recent report, Facebook introduced the concept of \textit{false amplifiers} that are characterized by ``coordinated activity by inauthentic accounts with the intent of manipulating political discussions''~\mbox{\citep{weedon2017information}}. The Facebook report introduces two new concepts associated with online information manipulation: \textit{coordination} and \textit{inauthenticity}. In literature, coordination is defined as an unexpected, suspicious, or exceptional similarity between a number of users~\mbox{\citep{nizzoli2021coordinated}}. In practice, it is often measured as the number of times two accounts behaved similarly, such as when they repeatedly retweet the same posts~\mbox{\citep{pacheco2020uncovering,nizzoli2021coordinated}}. Coordination among a certain number of users is considered to be a fundamental ingredient in successful manipulations, since it is necessary for the manipulation to obtain a significant outreach. On the contrary, the importance of inauthenticity -- that is, whether online users misrepresent themselves -- is still debated\footnote{\url{https://slate.com/technology/2020/07/coordinated-inauthentic-behavior-facebook-twitter.html}}. To this end, \textit{harmfulness}, rather than inauthenticity, appears to be more relevant for the analysis of online manipulations\footnote{\url{https://medium.com/1st-draft/how-to-improve-our-analysis-of-coordinated-inauthentic-behavior-a4ec62ce9bff}}.

Interestingly, coordination is an orthogonal concept with respect to the long-studied automation\footnote{\url{https://m.facebook.com/communitystandards/inauthentic_behavior/}}. In fact, recent results demonstrated limited correlations~\mbox{\citep{hristakieva2022spread}}, or no correlations at all~\mbox{\citep{nizzoli2021coordinated}}, between coordinated and automated accounts. In other words, coordination can be achieved without resorting to automation (i.e., social bots). Coordinated harassment by trolls and the supporting activity of online fandoms are two examples -- harmful and harmless, respectively -- of coordinated activities by human-operated accounts. For these reasons, the study of coordinated accounts and of their role in online information spreading can complement established studies on online information manipulation and can inform the development of new and effective decision support systems against these manipulations.

In this paper, we focus on the role of coordinated accounts in information spreading online -- a crucial, yet understudied, issue. In particular, we address the efficacy of their coordinated behavior in deceiving feed algorithms to maximize information outreach. We perform an analysis of a large dataset of 1.4 million tweets about the 2019 UK political elections. We find that coordinated and non-coordinated accounts significantly differ in terms of their retweeting activity and their positioning within information cascades. Additionally, their involvement in the cascades is much higher in empirical data that in randomized models. Finally, by analyzing the interactions between the two types of accounts, we uncover a saturation-like process due to which the maximal interaction between coordinated and non-coordinated accounts amounts to $\sim50\%$ of the cascade links, regardless of the penetration of coordinated accounts in the whole cascade.

The paper is structured as follows: Section~\mbox{\ref{sec:res_obj}} outlines our research objectives, the main findings and their significance. Section~\ref{sec: related work} presents related works on coordinated behavior and information spreading. Section~\ref{sec: data&methods} illustrates the dataset used for conducting the analysis and the employed measures and methods. Section~\ref{sec: results} details the results of the analysis. Section~\ref{sec:research implic} discusses the implications of our research. Finally, Section~\ref{sec:concl} reports the conclusions.

\section{Summary of research objectives, findings and significance}
\label{sec:res_obj}

\subsection{Research objectives and findings}
Our study on the differences between the interaction patterns of coordinated and non-coordinated accounts focuses on retweet dynamics, the primary way to spread information on Twitter~\mbox{\citep{firdaus2018retweet}}. Under this perspective, \textit{our overarching goal is that of investigating the role, impact and tactics of coordinated accounts in information spreading on Twitter}. In particular, we currently have limited knowledge of the peculiar traits, actions and tactics of coordinated accounts, except for their similarity in posting dynamics. Our work contributes to filling this gap by adding new insights over the limited existing knowledge.

We reach our overarching goal by achieving the following three research objectives:
\renewcommand\labelenumi{\bfseries{RO\theenumi:}}
\begin{enumerate}[leftmargin=3\parindent]
    \setcounter{enumi}{0}
    \item Describe the role of coordinated accounts in information spreading.
\end{enumerate}

In fulfillment of RO1, we identify three measures that significantly distinguish the activity of coordinated accounts from that of non-coordinated ones. Such measures are related to the position of coordinated accounts in the information cascade (i.e., the distance from the root), their action delay (i.e., the time required for retweeting a message) and their descendent counts (i.e., the number of users that a coordinated account is able to involve in the cascade). \textit{We find that, on average, coordinated accounts are closer to the root of the information cascade, are faster in retweeting original messages and are able to involve a higher number of users with respect to non-coordinated ones.}

\begin{enumerate}[leftmargin=3\parindent]
    \setcounter{enumi}{1}
    \item Assess the impact of coordinated accounts on information outreach and engagement of posts.
\end{enumerate}

To reach RO2, we carry out a thorough statistical analysis and we propose two novel indices. By mimicking a network dismantling process~\mbox{\citep{ren2019generalized}}, our indices provide a synthetic yet informative description of the impact that coordinated accounts have on information cascades. \textit{Our analysis reveals a systematic influence of coordinated accounts on the structure of the information cascades in which they participate, in that they are able to affect cascade size, number of links and depth.} Our findings are validated with appropriate null models.

\begin{enumerate}[leftmargin=3\parindent]
    \setcounter{enumi}{2}
    \item Investigate the tactics and targets of coordinated accounts.
\end{enumerate}

In RO3 we study the behavior of coordinated accounts in terms of targeted users. While we expected the amount of links between coordinated and non-coordinated accounts to grow proportionally with the number of coordinated accounts, \textit{we instead found a different situation where coordinated accounts are able to involve non-coordinated accounts in the cascade only up to a certain threshold that is $\sim50\%$ of the links.} This means that coordination requires a big effort in order to be effective and that massive coordinated actions are indistinguishable from less prominent ones in terms of involvement of non-coordinated accounts. Overall, our results contribute to shed light on the tactics of coordinated accounts and their influence on information diffusion.

\subsection{Significance}
Coordinated inauthentic and harmful behavior is an emerging problem on social media that has been linked to many other issues such as the spread of mis- and disinformation~\mbox{\citep{vargas2020detection}}. However, only limited knowledge, tools and results currently exist on coordinated behavior~\mbox{\citep{nizzoli2021coordinated}}. Not only does this lack of knowledge testify the existence of a scientific gap, but it also impairs the practical development of decision support systems (DSS) for supporting online platforms at maintaining safe and reliable social environments. Examples of such systems are those for assessing the veracity of online information~\mbox{\citep{lozano2020veracity}}. Our theoretical results on the role played by coordinated accounts in the online spread of information contribute to filling the current scientific gap. In addition, a deeper understanding of the characteristics and behavior of coordinated users, and of their tactics to tamper with the regular flow of information, also have practical implications since our results can be used as features in future machine/deep learning systems for detecting both coordinated accounts~\mbox{\citep{magelinski2021synchronized}} as well as inorganic or manipulated discussions~\mbox{\citep{tardelli2021detecting}}. Such systems will represent the next line of defense against coordinated propaganda and information manipulation, thus supporting both platform administrators and online users.

\section{Related works}
\label{sec: related work}
This section surveys relevant literature on online information manipulation and on online information spread.

\subsection{Coordinated inauthentic and harmful behavior}
Our present study is positioned within the growing body of work on coordinated inauthentic and harmful behavior, a recent stream of research that focuses on the collective behavior exhibited by the accounts involved in online information manipulations. The rationale for this body of work stems from the observation that online manipulations (e.g., disinformation campaigns) must reach a wide number of users to be successful. This mandates large and coordinated efforts so that the manipulation can obtain a significant outreach, exert influence, and thus have an impact~\mbox{\citep{tardelli2021detecting}}. Despite often appearing together, coordination, inauthenticity and harmfulness are distinct and orthogonal concepts~\mbox{\citep{nizzoli2021coordinated}}. For example, regarding coordination and inauthenticity, there can exist activists and other grassroots initiatives that are characterized by coordinated but authentic behaviors. Conversely, an ill-intentioned user might maneuver a single fake account, thus exhibiting inauthentic but uncoordinated behavior. Obviously, many other combinations of coordination, inauthenticity and harmfulness are possible.

To detect coordinated behaviors, existing works either leveraged similarities in user behaviors analyzed via network science frameworks~\mbox{\citep{hristakieva2022spread,nizzoli2021coordinated,pacheco2020uncovering,alassad2020combining,weber2020s,ng2021coordinating,schoch2022coordination}}, or temporal synchronicity between user actions~\mbox{\citep{magelinski2021synchronized,giglietto2020takes,zhang2021vigdet,sharma2021identifying}}. The former approach typically includes analytical steps such as the construction of a weighted user-similarity network, the filtering of such network, the detection of the different coordinated communities in the network, and the study of their extent and patterns of coordination~\mbox{\citep{weber2021amplifying}}. Contrarily to the positive results obtained for detecting coordinated behaviors, only limited results were achieved for distinguishing between authentic and inauthentic, or between harmless and harmful, coordination~\mbox{\citep{vargas2020detection}}. Some scholars worked around this issue by simply considering exceptional levels of coordination to be always indicative of malicious behaviors~\mbox{\citep{pacheco2020uncovering,giglietto2020takes}}. Others instead proposed to combine the analysis of coordination and propaganda, as a way to identify harmful coordinated communities~\mbox{\citep{hristakieva2022spread}}. Anyway, independently on the methods adopted for the analyses, the majority of existing works focused on detecting coordination, leaving the characterization of coordinated users and the understanding of their tactics and influence essentially unexplored.

\subsection{Information manipulation actors}
Still marginally related to our work is the body of research on social bots and trolls, two types of malicious actors involved in online information manipulations~\mbox{\citep{starbird2019disinformation}}. In fact, detecting and removing these accounts is still considered to be an effecting approach for curbing online manipulations~~\mbox{\citep{ferrara2016rise,fornacciari2018holistic}}. The main difference between social bots and trolls is that the former is -- at least partially -- automatically driven via software, whereas the latter is mostly human-driven. The possibility of performing actions automatically, without the need for human intervention, makes social bots the ideal information spreading and management tool, since they can easily perform many actions (e.g., resharing certain content) in a limited time span~\mbox{\citep{kang2021can}}. Based on this intuition, several studies investigated the role of social bots in the spread of disinformation and, more broadly, in the emergence of several ailments that affect our online social ecosystems (e.g., polarization, hate speech). To this end, existing results are controversial, with some studies reporting a significant contribution of social bots in the spread of low-credibility content~\citep{shao2018spread} and the extremization of online communities~\citep{stella2018bots}, while other studies only reported a marginal role~\citep{vosoughi2018spread}. Others specifically investigated the activity of bots in discussions about politics~\citep{woolley2016automating}, finance~\citep{tardelli2021detecting,mirtaheri2021identifying}, health~\citep{yuan2019examining} and entertainment~\citep{mendoza2020bots,mazza2019rtbust}, finding a much larger bot activity than that measured on average on social media. Regarding the characteristics and the detection of social bots, there exists a consensus that such accounts are becoming more and more sophisticated as a consequence of the technological tools (e.g., deepfake videos and profile pictures\footnote{\url{https://www.wired.com/story/facebook-removes-accounts-ai-generated-photos/}}) that are increasingly used to create credible fake online personas~\citep{cresci2018reaction}. These powerful computational means inevitably create increased challenges for the detection and removal of social bots, which also fueled criticism about the efficacy of the existing bot detection techniques~\citep{boneh2019relevant,rauchfleisch2020false}. For this reason, recent approaches to contrast online information manipulations avoid classifying the nature (i.e., automated or human-driven) of individual accounts, and rather focus on detecting and investigating suspicious patterns of coordination. The latter approach, also used in our work, is more novel and considered to be more effective than the former.

\subsection{Online information spreading}
Independently of the type and nature of the actors involved in online information manipulations, one of their aims is to alter spreading dynamics, to amplify their messages, and to extend their reach. In particular, focusing on Twitter, the simplest action that can be used to spread messages to the Twittersphere is the \emph{retweet}~\citep{firdaus2018retweet,mazza2019rtbust}. Retweeting a message means forwarding it to the user's followers network. Despite the simplicity of the process behind a retweet cascade graph, the free Twitter API service does not provide full information about the spreading process. Instead, it only gives information about each retweeter and the author of the original tweet. In other words, the actual path followed by an original tweet to reach its audience is unknown and must be inferred/reconstructed~\citep{vosoughi2017rumor}. Initial works \citep{DBLP:conf/www/YangHZSG12,DBLP:journals/corr/zaman,DBLP:conf/cikm/CaoSCOC17} derived the retweet propagation paths assuming that the user's ``screen name'' reported in the tweet text, such as ``\textit{RT@ screen name}'' indicates the user from whom the current user read the message. However, this assumption is mostly inaccurate \citep{DBLP:conf/sitis/CazabetPTT13} and information cascades that merge temporal data with social network information are generally considered to be more reliable. In such a vein, \citep{DBLP:conf/www/TaxidouF14} proposed a model analyzing four different options based on followed accounts to derive the possible retweet graph. Since there is no ground truth to compare the possible cascade options, the authors evaluated the options computing several metrics. Other works considered additional information to derive the retweets' propagation path such as text and topic similarity features \citep{yang2010predicting} or location information \citep{DBLP:journals/tkde/WuCZCLM20,DBLP:conf/worldcist/RodriguesCIPS16}. Few studies integrated the impact of social relationships, measured in terms of reciprocal interactions \citep{DBLP:journals/tkde/WuCZCLM20,DBLP:conf/cikm/CaoSCOC17,zola2020interaction}, in the analysis of retweeting dynamics. In particular, the recent work of \citep{zola2020interaction} introduced two models aimed at measuring the strength of interactions among users to build weighted retweet cascade graphs.

In this work, we are interested in studying the role of coordinated users in information cascades. For reconstructing the cascades we adopt the state-of-the-art method proposed in~\mbox{\citep{de2015towards}} that leverages the follower-following network of the users that participate to an information cascade, thus merging information deriving from users' interests and social dynamics.

\section{Data and Methods}
\label{sec: data&methods}

\subsection{Data}
\label{subsec:data}
Our dataset for this study is based on a large collection of tweets related to the online debate about the 2019 United Kingdom general election\footnote{\url{https://en.wikipedia.org/wiki/2019_United_Kingdom_general_election}}. By leveraging Twitter Streaming APIs, we collected one month of data: from 12 November to 12 December 2019 (election day). Our data collection strategy was based both on election-related hashtags and on influential political accounts. Regarding hashtags, we collected all tweets mentioning at least one hashtag from the list shown in Table~\ref{tab:dataset-hashtags}. This list contains both polarized (i.e., labour or conservatives) hashtags, as well as neutral ones. Furthermore, we also collected all tweets published by the two parties' official accounts and by their leaders, shown in Table~\ref{tab:dataset-accounts}, together with all the interactions (i.e., retweets and replies) they received. This data collection process allowed us to gather 11,264,820 tweets published by 1,179,659 distinct users. This dataset is publicly available for research purposes\footnote{\url{https://doi.org/10.5281/zenodo.4647893}}.

\begin{table}[h]
    \begin{minipage}{.45\linewidth}%
	\small
	\centering
	\begin{tabular}{@{}lrrr@{}}
		\toprule
		\textbf{hashtag} & \textit{users} & \textit{tweets} \\
		\midrule
		\textsf{\#GE2019}	                & 436,356			& 2,640,966			\\
		\textsf{\#GeneralElection19}	    & 104,616			& 274,095			\\
		\textsf{\#GeneralElection2019}	    & 240,712			& 783,805			\\
		\textsf{\#VoteLabour}	            & 201,774			& 917,936			\\
		\textsf{\#VoteLabour2019}	        & 55,703			& 265,899			\\
		\textsf{\#ForTheMany}	            & 17,859			& 35,621			\\
		\textsf{\#ForTheManyNotTheFew}	    & 22,966			& 40,116			\\
		\textsf{\#ChangeIsComing}	        & 8,170			    & 13,381            \\
		\textsf{\#RealChange}	            & 78,285			& 274254			\\
		\textsf{\#VoteConservative}	        & 52,642			& 238,647			\\
		\textsf{\#VoteConservative2019}	    & 13,513			& 34,195			\\
		\textsf{\#BackBoris}	            & 36,725			& 157,434			\\
		\textsf{\#GetBrexitDone}	        & 46,429			& 168,911			\\
		\midrule
		\textbf{total}	                    & 668,312			& 4,983,499		        \\
		\bottomrule
	\end{tabular}
	\caption{Statistics about data collected via hashtags.}
	\label{tab:dataset-hashtags}

    \end{minipage}%
    \hspace{0.1\linewidth}%
    \begin{minipage}{.45\linewidth}%
	\small
	\centering
	\begin{tabular}{@{}lrrrr@{}}
		\toprule
		&&& \multicolumn{2}{c}{\textbf{interactions}}\\
		\cmidrule{4-5}
		\textbf{account} & \textit{tweets} && \textit{retweets} & \textit{replies}\\
		\midrule
		\textsf{@jeremycorbyn}  & 788   && 1,759,823    & 414,158   \\
		\textsf{@UKLabour}      & 1,002 && 325,219      & 79,932    \\
		\textsf{@BorisJohnson}  & 454   && 284,544      & 382,237   \\
		\textsf{@Conservatives} & 1,398 && 151,913      & 169,736   \\
		\midrule
		\textbf{total}          & 3,642	&& 2,521,499    & 1,046,063 \\
		\bottomrule
	\end{tabular}
	\caption{Statistics about data collected from accounts.}
	\label{tab:dataset-accounts}

    \end{minipage}%
\end{table}

Since we are particularly interested in coordinated behaviors and their effects on information diffusion, we subsequently derived coordination scores for each user in our dataset by applying the state-of-the-art method proposed in~\mbox{\citep{nizzoli2021coordinated}}. This method, as well as other similar ones~\mbox{\citep{pacheco2020uncovering}}, measures coordination between accounts by analyzing the sequence of retweets of each account and by subsequently computing pairwise similarities between all accounts. In particular, we associated each account to a TF-IDF weighted vector of the tweet IDs retweeted by that account. The TF-IDF weighting reduces the relevance of highly popular tweets, that are likely to receive many retweets, in favor of unpopular ones, whose retweets might be indicative of suspicious behaviors~\mbox{\citep{mazza2019rtbust}}. Then, we computed user similarities as the cosine similarity of account vectors. The similarity scores between all couples of accounts are used to build a user-similarity network whose edges are ranked by non-increasing weight. In our work, we retained the top 1\% of the edges with the highest similarity scores among all those in our dataset. Following previous work~\mbox{\citep{pacheco2020uncovering}}, the nodes connected by such edges are considered to be coordinated accounts. The remaining accounts in our dataset are considered to be non-coordinated.

Finally, we filtered our initial dataset by only retaining the 49,331 tweets that were retweeted at least once by the accounts classified as coordinated and whose author is not a coordinated user. For each such tweet in our sample, we constructed its retweet-based information cascade by following the method proposed in \citep{de2015towards}, as thoroughly detailed in the next subsection. Descriptive statistics about the retweeters (i.e., users who retweeted at least one post) are reported in \cref{tab: descrizione dati}. 

\begin{table}[h]
    \centering
    \begin{tabular}{ccc}
    \toprule
       \textbf{average retweeters}  & \textbf{number of unique} & \textbf{number of coordinated} \\
        \textbf{per cascade} & \textbf{retweeters} & \textbf{retweeters} \\
    \midrule
    16.38 & 205,277 & 1,192 \\
    \bottomrule
    \end{tabular}
    \caption{Summary features of retweeters. Among the 205,277 retweeters, 1,192 are labelled as coordinated.}
    \label{tab: descrizione dati}
\end{table}

\subsection{Retweet information cascades}
\label{subsec: retweet cascade graph}
To investigate the role of coordinated accounts in spreading information, we first need to construct the information cascades themselves. The easiest way to spread a message on Twitter is through the retweet action. Thus, on Twitter, information cascades can be studied by leveraging retweet graphs. The Twitter API service provides detailed information about each tweet (e.g., engagement values, user information), but it does not provide enough information to automatically reconstruct retweet graphs. Therefore, given a user who retweeted a tweet, it is not possible to determine with certainty from which other node of the graph the user saw and retweeted the tweet. To solve this limitation and to infer the full propagation structure of a tweet, we follow the well-known approach proposed in~\mbox{\citep{de2015towards}} that derives the retweet graph based on the social network of each user and on the retweet timestamps. In detail, given a retweet $t_x$ by user $u_x$, of an original tweet $t_0$ by user $u_0$, the retweeting user $u_x$ might have either retweeted the original tweet $t_0$, or another retweet $t_i$ of $t_0$ posted by one of his/her friends ($u_i$) before him/her. Thus, $0 \leq i < x$. To identify the user from which $u_x$ retweeted,~\mbox{\citep{de2015towards}} proposes to select the one that posted/retweeted $t_0$ most recently before $u_x$, thus leveraging the publication timestamp of the original tweet $t_0$ and those of all other retweets $t_i$ of $t_0$ published before $t_x$. Notably, information about the social network and the timestamps are available from the Twitter APIs. The simplicity of this method for reconstructing retweet graphs made it one of the most widely used in recent literature~\mbox{\citep{vosoughi2017rumor}}.

Once reconstructed, the retweet graph $C$ is defined as a directed graph $C=(V,E)$, where each node $u \in V$ represents a user $u$ and an edge $(u,v) \in E$ represents a link from user $u$ to user $v$. The graph $C$ is not necessarily weakly connected since some groups of interconnected nodes or singletons could be disconnected from the largest connected component of $C$ -- i.e., from the largest subgraph of $C$ made up of interconnected nodes. The largest connected component of $C$, called $C'=(V',E')$, is a directed tree having as root node $r$ the user $u_0$ -- that is, the original author of the tweet $t_0$ retweeted by all other nodes in $V$. The subset of $V$ made up of nodes that do not belong to $V
'$ is called $S$. Users in $S$ are those who retweeted the post of $r$ by reading it either from the Twitter trends or by using the keyword search function.
Obviously, if $S$ is an empty set then $C \equiv C'$.

Once obtained the retweet graph, it is possible to compute several global metrics, listed below.
\begin{itemize}
    \item Cascade size: the number of nodes in the information cascade $C$. The cascade size is denoted as $s = |C| = |C'| + |S|$,  where $|\cdot|$ is the cardinality of the considered set. Accordingly, the size of $C'$ is denoted as $s'=|C'|$.
    \item Cascade edges: the number of edges in the information cascade $C$. The cascade edges are denoted by their count $m = |E|$. Accordingly, the cascade edges in $C'$ are denoted as $m'=|C'|-1=s'-1$, whereas the latter equality derives from the definition of tree graph.
    \item Cascade height: the length of the longest path going from $r$ to any other node in $C'$. More intuitively, the cascade height $h'$ can be thought of as the number of levels in the cascade assuming that the root node is at level 0, the nodes at distance 1 from the root are at level 1, and so on. The cascade height  $h'$ measures the extent to which a tweet was able to reach users far from the root's ego network. 
\end{itemize}

Alongside macro measures related to $C$ and $C'$, we can introduce other measures at node-level and edge-level. The nodal measures that we take into account are:
\begin{itemize}
    \item Node positioning level: given the node $u \in V'$, its positioning level $l'_u$ is an integer number that quantifies its distance from the root $r$.
    \item Node action delay: the action delay $a_u^\triangleleft$ denotes the number of minutes passed since the original tweet posting time and the node $u$ retweet time.
    \item Node descendants: the number of links outgoing from node $u$. This measure is also known as node out-degree $k_u^{out}$.
\end{itemize}

\begin{figure}[h]
    \centering
    \includegraphics[scale=0.4]{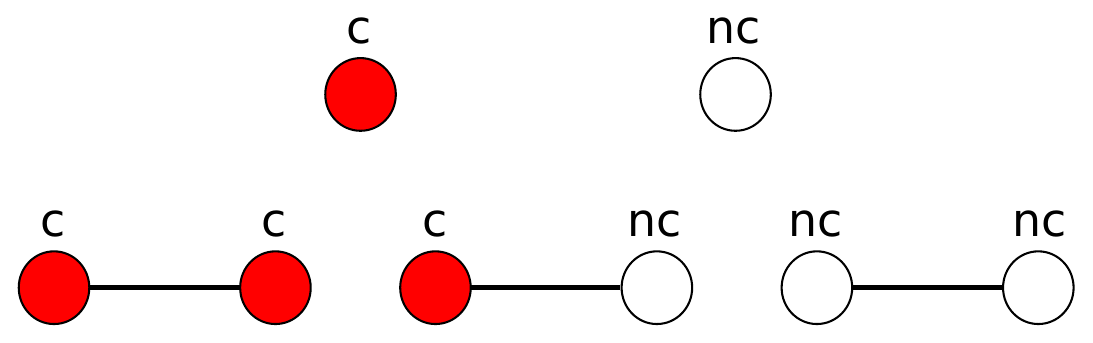}
    \caption{Node and edge types. A red node is a coordinated one (c) while a white node is a non-coordinated node (nc). Consequently, there are three types of edges: (c -- c), (c -- nc) and (nc -- nc). Node counts are refereed to as $s_c$ and $s_n$, while edge counts are referred to as $m_{cc}$, $m_{cn}$ and $m_{nn}$.}
    \label{fig:node_edge_classification}
\end{figure}

The edge-level measures can be obtained using counts of edge types deriving from the node classification into coordinated and non-coordinated ones. Similar binary classifications of nodes  and edges have been used also in~\citep{park2007distribution, cinelli2020network}. Given the information cascade $C=(V,E)$, the nodes in $V$ that define the cascade size $s$ can be divided into two subsets, the set of coordinated ones of cardinality $s_c$ and the set of non-coordinated ones of cardinality $s_n$. Obviously, $s=s_c+s_n$. We define coordinated to coordinated edges the set of edges connecting couples of coordinated nodes. The cardinality of such a set is $m_{cc}$. Similarly, we define coordinated and non-coordinated edges, the set of edges between couples of coordinated to non-coordinated nodes. The cardinality of such a set is $m_{cn}$ and $m_{cn}=m_{nc}$. Finally, we define non-coordinated to non-coordinated edges, the set of edges connecting couples of non-coordinated nodes. The cardinality of such a set is $m_{nn}$. Obviously, $m=m_{cc}+m_{cn}+m_{nn}$. An example of the classification of nodes and edges is displayed in Figure~\ref{fig:node_edge_classification}. Accordingly, the tree represented in Figure~\ref{fig:cir e ctncir} has $s_c=7$, $s_n=16$, $m_{cc}=4$, $m_{cn}=8$ and $m_{nn}=10$.

\subsection{Accounts' Infectivity ratios}
\label{subsec: indici per spread info}

Given the features assigned for each node $u \in V$, we introduce two novel metrics to measure the impact of coordinated users in spreading information along the cascade. The two metrics are respectively called coordinated accounts infectivity ratio (C$_{IR}$) and coordinated to non-coordinated accounts infectivity ratio (CtNC$_{IR}$).

\begin{figure*}
    \centering
    \includegraphics[scale=0.32]{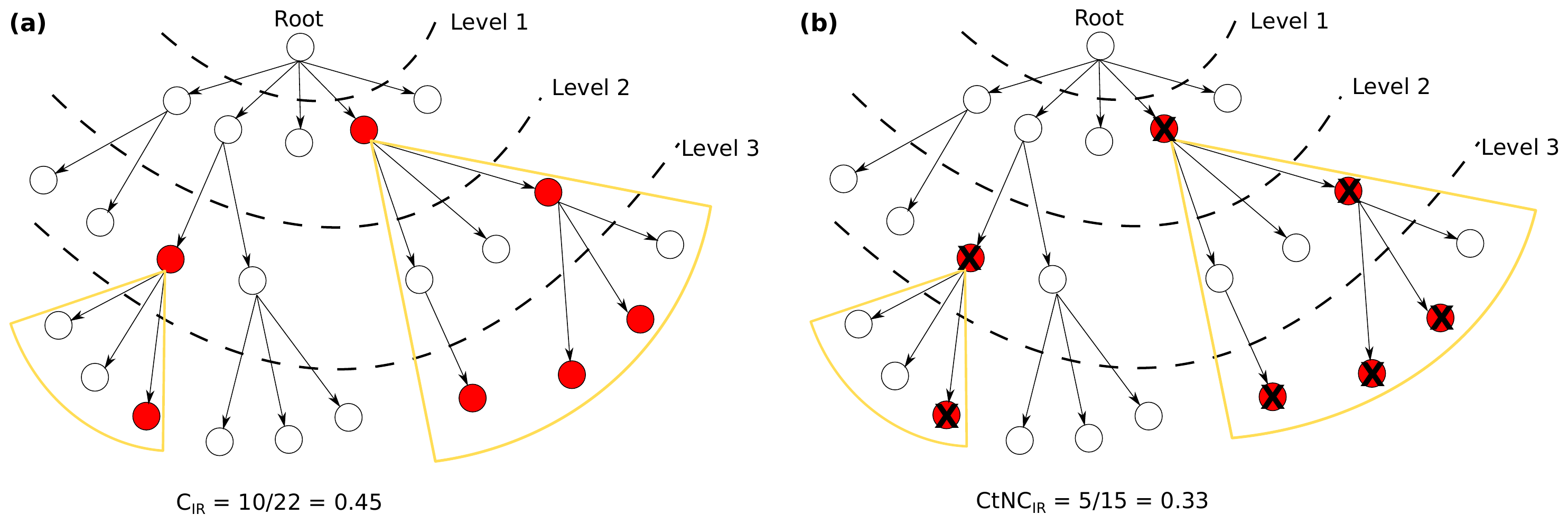}
    \caption{Example of C$_{IR}$ and CtNC$_{IR}$ metrics}
    \label{fig:cir e ctncir}
\end{figure*}

In particular, the C$_{IR}$ counts what proportion of nodes in the cascade depends, in a topological sense, on coordinated nodes. In other words, C$_{IR}$ counts the proportion of nodes no longer reachable from the root $r$ of $C'$, after removing the coordinated nodes from the cascade (counting the coordinated nodes as well). The CtNC$_{IR}$ follows the same rationale of C$_{IR}$, but it excludes coordinated nodes from the count. As such, it is more representative of the effect that removing coordinated accounts has on non-coordinated ones (i.e., unaware social media users).
An example of both measures is reported in \cref{fig:cir e ctncir}.

In the following we provide the procedure to compute the two aforementioned metrics. 
\renewcommand\labelenumi{\theenumi.}
\begin{enumerate}
\item Consider a cascade $C$ rooted in the node $r$.
\item Sort the coordinated nodes by increasing distance from the root $r$ (nodes disconnected from the largest connected component and with no incoming links are assumed to be at distance 1 from the root). 
\item Choose the induced subgraph having as root node the coordinated node at the minimum distance from the root $r$. 
\item Count the number of nodes in the subgraph excluding its root and store such a count in a vector called $v_C$ (C$_{IR}$ case).
\item Count the number of non-coordinated nodes in the subgraph excluding its root and store such a count in a vector called $v_{CtNC}$ (CtNC$_{IR}$ case).
\item Remove from $C$ all the nodes in the subgraph.
\item Iterate Step 3 to Step 6 until there are no more coordinated nodes in $C$.
\item Sum over the elements of $v_C$ and divide the sum by $s-1$ to obtain C$_{IR}$.
\item Sum over the elements of $v_{CtNC}$ and divide the sum by $s-s_c-1$ to obtain CtNC$_{IR}$.
\end{enumerate}

\section{Results}
\label{sec: results}
Based on the definitions and the metrics previously introduced, in this section we report and discuss the results of our analyses.

\subsection{Coordination-based difference in information spreading}
\label{subsec: result Rq.1}
In order to understand if coordinated accounts differ from non-coordinated ones in their retweeting behavior, we perform a statistical analysis of the node-level measures introduced in Section~\ref{sec: data&methods}.

\begin{figure}[h]
    \centering
    \includegraphics[scale=0.55]{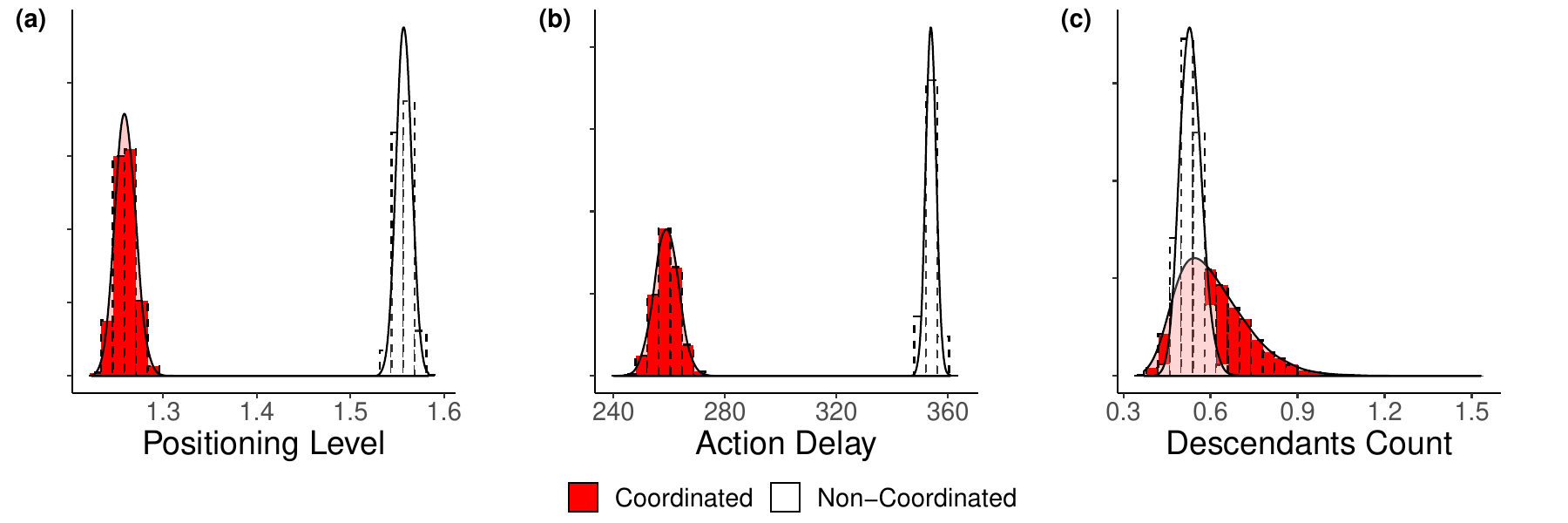}
    \caption{Distribution of node-level features for coordinated and non-coordinated accounts.}
    \label{fig:boot}
\end{figure}

Figure \ref{fig:boot} reports the kernel density estimations of the distributions of nodes' positioning level, action delay, and descendants count, respectively. Such densities are obtained using average values computed across 50,000 bootstrap samples on both coordinated and non-coordinated sets of nodes (summary statistics and differences in the distributions are reported in Table~\ref{tab:bootstrap_stats}). The reason behind using bootstrap samples is that of smoothing out the disproportion in the number of coordinated and non-coordinated accounts (as also reported in Table~\ref{tab: descrizione dati}). Figure \ref{fig:boot}(a) displays a small but significant difference for what concerns the level occupied by coordinated and non-coordinated accounts in the retweet cascade. Indeed, the former group occupies, on average, higher levels meaning that coordinated accounts tend to be closer to the root of the cascade. Accordingly, Figure~\ref{fig:boot}(b) shows that coordinated accounts are much faster in retweeting information, as shown by the significantly lower action delay. 

Analyzing \cref{fig:boot}(c), it may be noticed that the two sets of users have a similar average number of descendants but, overall, we find that the two distributions differ. In particular, coordinated accounts display a right-skewed distribution of descendent count, thus being able to attract a higher number of descendants nodes, that is, the observed difference is mostly due to the right tail that characterize the descendent count distribution of coordinated accounts.
For all the considered measures we find a statistically significant difference between the two sets of users. In summary, our results suggest a remarkable difference in the way information is spread by coordinated and non-coordinated accounts.

\begin{table}[!htpb]
\centering
\begin{adjustbox}{width=\textwidth}
\begin{tabular}{l|rrrrr|rrrrr|c}
\toprule
\multicolumn{1}{c|}{\multirow{2}{*}{\textbf{}}} & \multicolumn{5}{c|}{\textbf{Coordinated}} & \multicolumn{5}{c|}{\textbf{Non-Coordinated}} & \multirow{2}{*}{\textbf{\begin{tabular}[c]{@{}c@{}}Wilcoxon Test \\ Significance\end{tabular}}} \\ \cline{2-11}
\multicolumn{1}{c|}{} & \textbf{Mean} & \textbf{St Dev} & \textbf{Median} & \textbf{min} & \textbf{max} & \textbf{Mean} & \textbf{St Dev} & \textbf{Median} & \textbf{min} & \textbf{max} &  \\
\hline
Positioning Level & 1.26 & 0.01 & 1.26 & 1.22 & 1.31 & 1.56 & 0.01 & 1.56 & 1.53 & 1.58 & * \\
Action Delay (minutes) & 258.91 & 4.30 & 258.94 & 240.02 & 274.61 & 353.89 & 1.79 & 353.87 & 347.71 & 360.84 & * \\
Descendent Count & 0.61 & 0.12 & 0.59 & 0.37 & 1.53 & 0.53 & 0.04 & 0.53 & 0.42 & 0.69 & *\\
\bottomrule
\end{tabular}
\end{adjustbox}

\caption{Descriptive statistics of bootstrap experiments. The differences in the distributions of the considered measures are evaluated by means of the non-parametric Wilcoxon signed-rank test.}
\label{tab:bootstrap_stats}
\end{table}

Our results also provide useful information about the difference between coordinated and merely active accounts. We remark that all accounts considered in our work were extremely active during the data collection period, since we specifically focused our analyses on superspreaders -- that is, the top-1\% of users who shared the most retweets in the run up to the 2019 UK general election. However, in spite of the fact that all users in our dataset were very active accounts, not all of them were coordinated, as demonstrated by our analyses. In other words, being very active does not necessarily imply being coordinated. In terms of our analyses and results, this means that there are many accounts that are very active and that often lay near the root of the information cascades to which they participate, but that are not coordinated. In turn, this suggests that coordination and ``activeness'' are two related, yet distinct, phenomena, which also motivates the recent interest in studying online coordination. More specifically, coordination is a stronger condition than simple ``activeness'', since it mandates that an account is active, but also that its activity is done in a somewhat synchronized way with the activity of other accounts.

\subsection{Influence of coordinated accounts on message outreach}
\label{subsec: result Rq.2}
Here, we investigate the relationship between the number of coordinated accounts involved in a message diffusion process and the message outreach (i.e., the magnitude of the information cascade generated by the message). Therefore, we analyze the structure of the retweet cascades by means of three measures, the cascade size, the cascade edges and the cascade height. Such measures are considered as covariates with respect to the incidence (i.e., the proportion) of coordinated accounts in the retweet cascade.  

\begin{figure}[ht]
    \centering
    \includegraphics[scale=0.5]{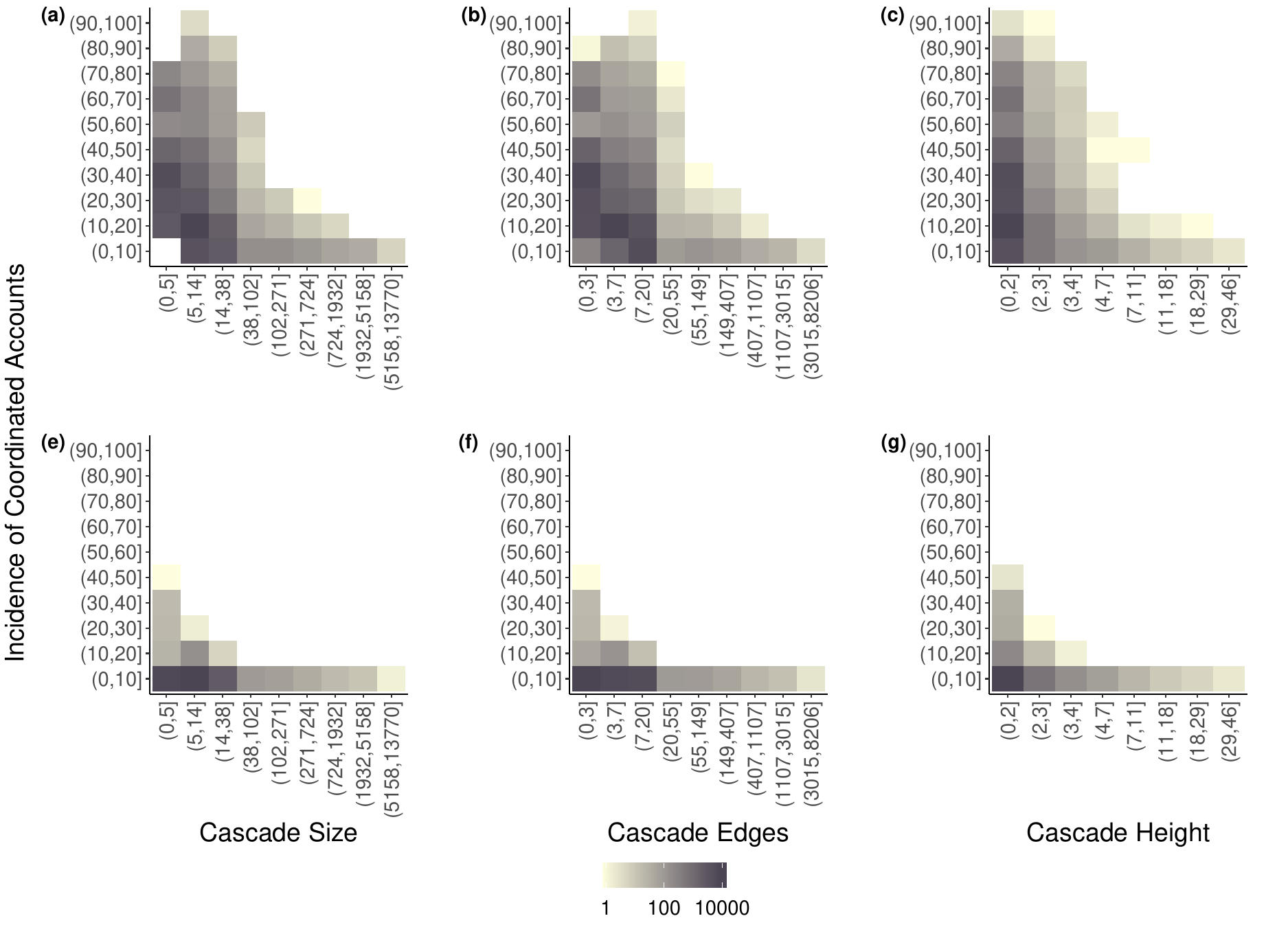}
    \caption{Incidence of Coordinated Account with respect to Size (left column), Edges (center column) and Height (right column) of retweets information cascades obtained for empirical data (top row) and null models (bottom row).}
    \label{fig:cascades_global}
\end{figure}

Qualitative results, reported in Figure~\ref{fig:cascades_global}, show structural features of the cascades (in terms of nodes, edges and height) on the \textit{x} axis and the incidence of coordinated accounts (i.e., the fraction between coordinated nodes over all nodes in a cascade) on the \textit{y} axis. Colors indicate the number of cascades lying in the plots' different areas, highlighting that large cascades are rare with respect to smaller ones. This first finding from our dataset supports previous results about information diffusion on Twitter~\citep{mendoza2020bots,zang2017quantifying}. Next, by comparing the actual placement of coordinated accounts (Figure~\ref{fig:cascades_global} (a), (b) and (c)) with that obtained via a null model that randomizes the node labels while fixing their proportions (Figure~\ref{fig:cascades_global} (e), (f) and (g)), we find that coordinated accounts are present mostly in small cascades. In other words, the bigger the viral process, the smaller the likelihood of their effect. This result suggests that the use of coordinated accounts to alter information spreading works essentially only for the initial engagement of users within a cascade, and not for the later and deeper stages of the diffusion process that are instead independent on the coordinated accounts and driven by spontaneous and unorganized user activity. This hypothesis is also supported and reinforced by our previous results on the placement of coordinated accounts within a cascade. Indeed, in Figure \ref{fig:boot}(a) we showed that coordinated accounts tend to reshare messages in the early stages of diffusion, given that they lay relatively near to the root of the cascades.

The findings reported in this section are fascinating in the context of the tactics used by coordinated accounts for altering information diffusion. In fact, while several previous studies found that automated and coordinated accounts were used for re-sharing messages~\citep{mazza2019rtbust,mendoza2020bots,giatsoglou2015retweeting,vo2017revealing,gupta2019malreg}, little was known about how such accounts affected the information diffusion process (which is instead seen from our results). When interpreting these results, care should also be taken about the expected impact of coordinated behaviors as obtained by the implemented null model. In fact, panels (e) to (g) of Figure~\ref{fig:cascades_global} show a relatively low incidence of coordinated accounts that might seem to indicate a rather weak or ineffective effort. On the contrary, however, we remark that the typical goal of coordinated users is to convey their message to the largest possible number of unaware (and uncoordinated) users in the network. As such, positive results are related to large cascades having a relatively low incidence of coordinated users, rather than those with an extremely high incidence of coordinated users. In fact, the latter would mean that the message almost did not reach any user, apart from those already involved in the manipulation.

Finally, our results' interpretation remarks an important challenge: the need to differentiate between authentic and inauthentic activities. In fact, large cascades with a low incidence of coordinated accounts could result both from a successful manipulation activity (i.e., inauthentic behavior) and from an authentic grassroots initiative (possibly also involving a minority of coordinated accounts). Distinguishing between these two forms of online activity currently represents one of the important open issues in this field~\citep{nizzoli2021coordinated, vargas2020detection}.

\subsection{Testing influence and targeting tactics of coordinated accounts}
\label{subsec: result Rq.3}
To better understand the dynamics of influence and targeting perpetrated by coordinated accounts, we analyze their infectivity ratios (i.e., their ability to involve other accounts in the spreading process) as well as the abundance of links with their targets in information cascades.

\paragraph{Infectivity ratios}
\label{par: infectivity ratios}

We consider the two infectivity ratios, C$_{IR}$ and CtNC$_{IR}$, introduced in Section~\ref{subsec: indici per spread info}. Such ratios count the proportion of nodes in the cascade that descend, directly or indirectly, from coordinated accounts. More precisely, the two measures are aimed at understanding which portion of the nodes in the cascade stems from the action of coordinated account (C$_{IR}$) and which portion of non-coordinated accounts in the cascade stems from the action of coordinated accounts (CtNC$_{IR}$).

\begin{figure}[h]
    \centering
    \includegraphics[scale=0.5]{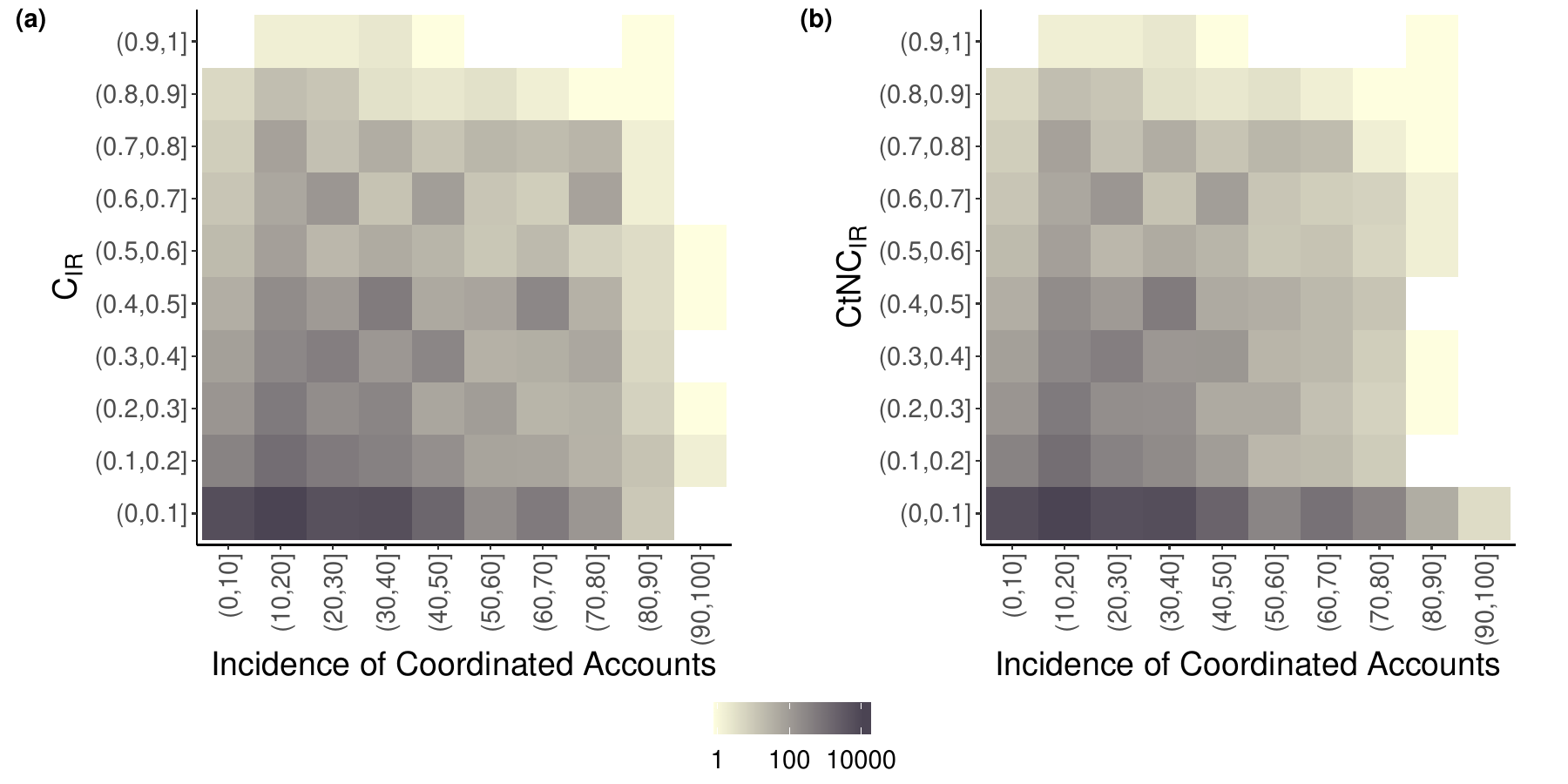}
    \caption{Infectivity ratios. Panel (a) displays the joint distribution of the Coordinated Infectivity Ratio (C$_{IR}$) and the incidence of coordinated accounts. Panel (b) displays the joint distribution of the Coordinated to Non-Coordinated Infectivity Ratio (CtNC$_{IR}$) and the incidence of coordinated accounts.}
    \label{fig:infectivity_ratios}
\end{figure}

Panels (a) and (b) of Figure \ref{fig:infectivity_ratios} show the coordinated accounts infectivity ratios by considering either all nodes (C$_{IR}$) or only non-coordinated ones (CtNC$_{IR}$). Both panels display a very scattered dynamic, possibly more focused on low values of the index in the C$_{IR}$ case. In general, coordinated accounts do not seem to pursue an effective placement strategy, despite having a higher average position level (as shown in Section~\ref{subsec: result Rq.1}). Their removal, indeed, doesn't systematically guarantee the structural collapse of the information cascade.
Furthermore, the positioning of coordinated accounts is rarely effective regardless the cascade size.

\begin{figure}[h]
    \centering
    \includegraphics[scale = 0.6]{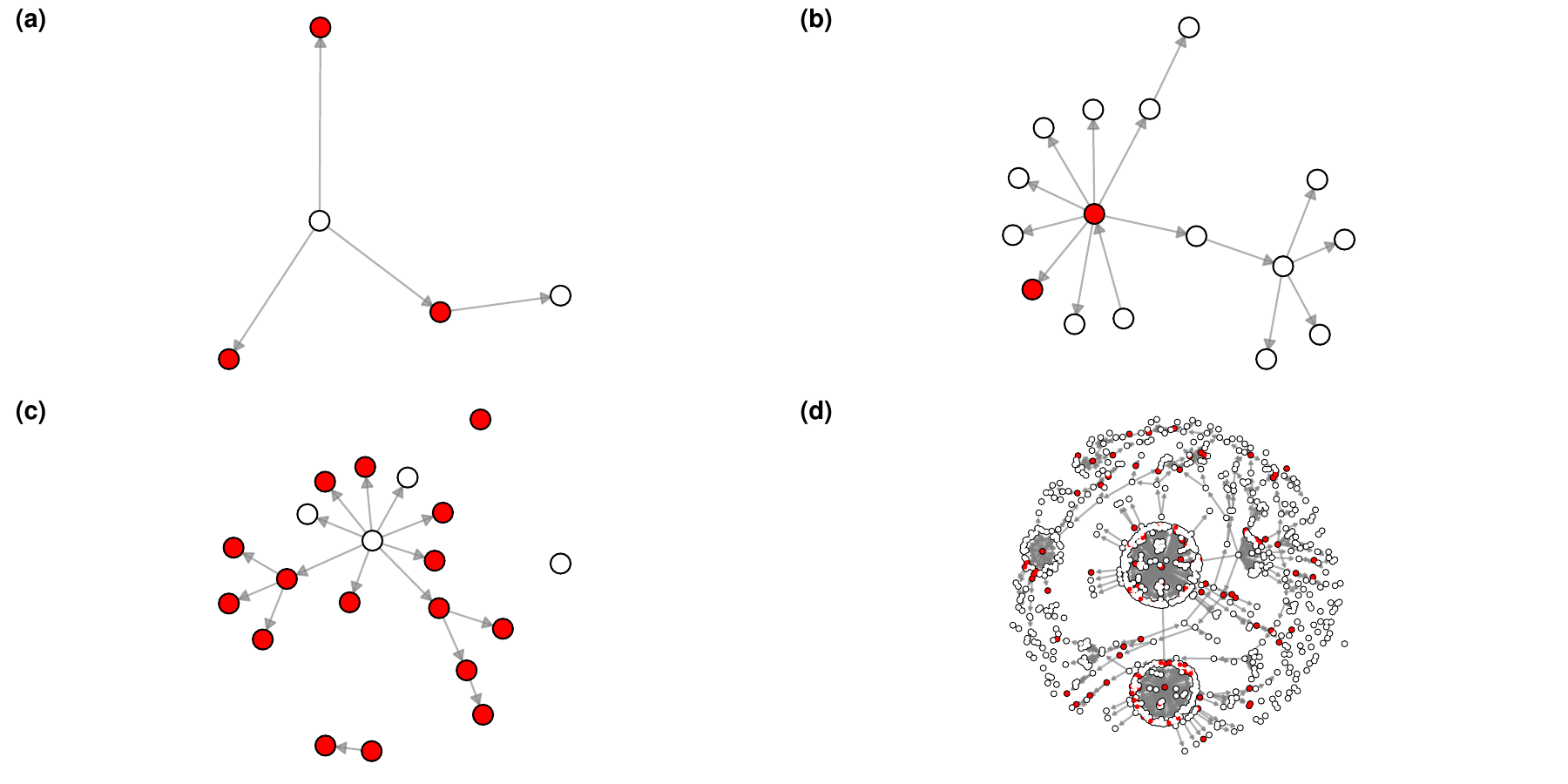}
    \caption{Examples of cascades displaying peculiar cases of C$_{IR}$ and CtNC$_{IR}$ indexes. Coordinated nodes are depicted in red while non-coordinated nodes are depicted in white.}
    \label{fig:esempi_cascate}
\end{figure}

Despite the scattered results of the infectivity ratios, a joint investigation of the distributions of the two indices provides interesting results. In \cref{fig:esempi_cascate} we report some notable examples of cascades from our dataset. In figure, white nodes denote non-coordinated accounts while red nodes denote coordinated ones.

\begin{table}[!htpb]
\centering
\label{Tab: cascate esempi}
\begin{tabular}{lllll}
\toprule
\multirow{2}{*}{Cascade} & \multirow{2}{*}{\begin{tabular}[c]{@{}l@{}}Nodes\\ Count\end{tabular}} & \multirow{2}{*}{\begin{tabular}[c]{@{}l@{}}Coordinated\\  Count\end{tabular}} & \multirow{2}{*}{C$_{IR}$} & \multirow{2}{*}{CtNC$_{IR}$} \\
 &  &  &  &  \\
\hline
a & 5 & 3 & 0.25 & 1 \\
b & 16 & 2 & 0.93 & 1 \\
c & 20 & 16 & 0.37 & 0 \\
d & 911 & 108 & 0.82 & 0.82\\
\bottomrule
\end{tabular}
\caption{Cascades and Infection rates examples.}
\end{table}

Details of \cref{fig:esempi_cascate} are reported in \cref{Tab: cascate esempi} which shows four different relations between the two infection indices, resulting in the following observations:
\begin{itemize}
    \item If C$_{IR} \longrightarrow0$ \& CtNC$_{IR} \longrightarrow0$: the cascade depends only on the root and coordinated nodes do not interfere in information spreading;
    \item If C$_{IR} \longrightarrow 1$ \& CtNC$_{IR} \longrightarrow 1$: coordinated accounts play a key role in spreading information to non-coordinated accounts;
    \item If C$_{IR} \longrightarrow 0$ \& CtNC$_{IR} \longrightarrow 1$: coordinated nodes are likely to be located next to the root of the tree and have few non-coordinated descendants. Such a case could be due mainly to finite size effects thus happening only in small cascades.
    \item If C$_{IR} \longrightarrow 1$ \& CtNC$_{IR} \longrightarrow 0$: the coordinated nodes have descendants but those descendants are coordinated as well. Thus, the impact of coordinated accounts on unaware (i.e., non-coordinated) users is negligible.
\end{itemize}

This taxonomy and the examples in~\mbox{\cref{fig:esempi_cascate}} reveal interesting phenomena and highlight the usefulness of the two indices in providing a deeper understanding about the role and the impact of coordinated accounts. In addition to their theoretical value, these results also have practical implications. For instance, the C$_{IR}$ and CtNC$_{IR}$ indices allow distinguishing between coordinated manipulations that managed to influence unaware (non-coordinated) users with those who did not. The former are particularly worthy of attention since, by influencing unaware users, they are likely to have serious consequences. When used in conjunction, the two indices would thus allow prioritizing moderation interventions by platform administrators where they are most needed.

\paragraph{Link classification results}
\label{par:link_class results}

\begin{figure}[h]
    \centering
    \includegraphics[scale=0.4]{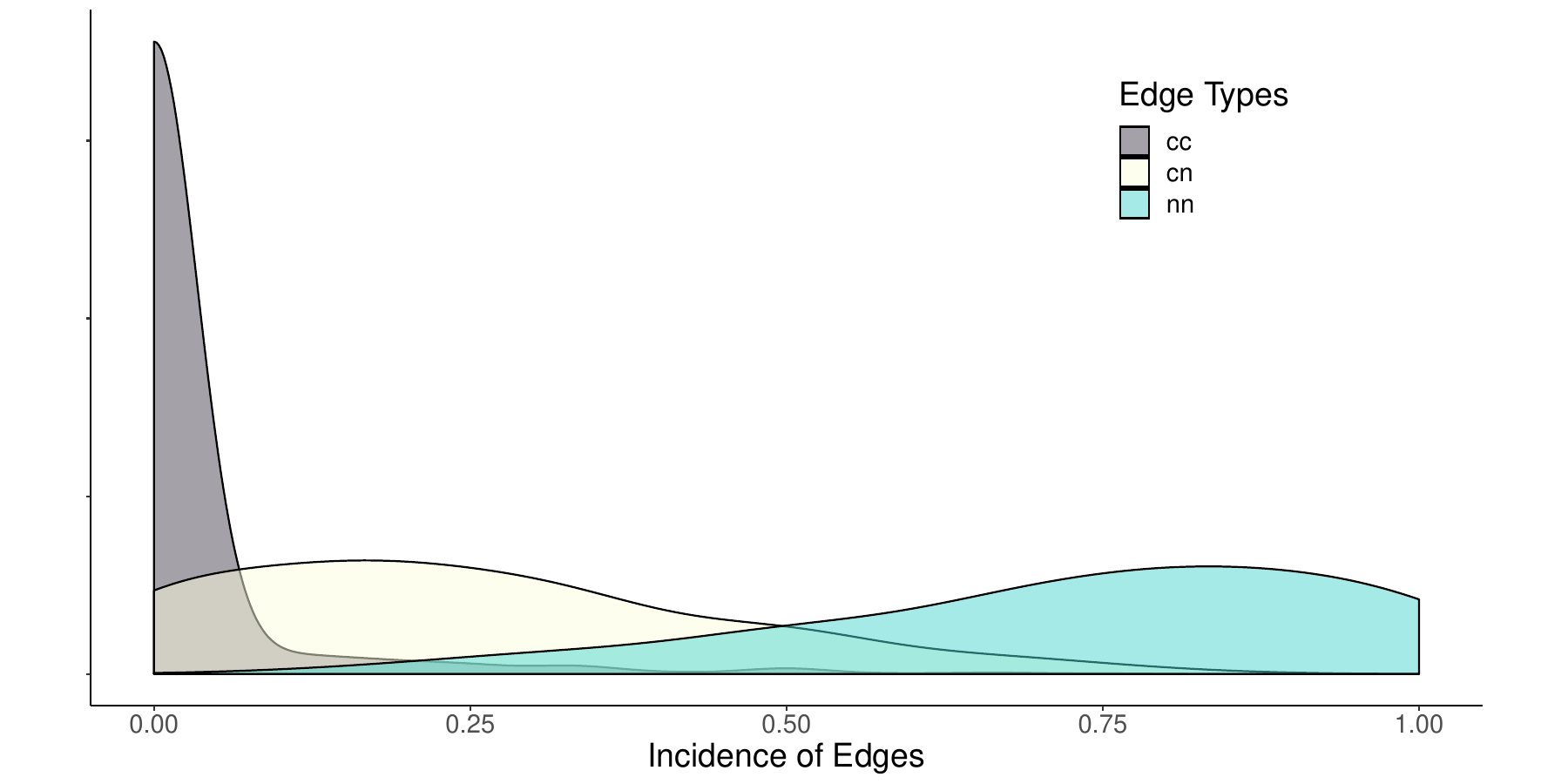}
    \caption{Kernel density estimates for the distribution of incidence of edge types.  Coordinated to coordinated edges are referred to as cc; coordinated to non-coordinated edges are referred to as cn; and non-coordinated to non-coordinated edges are referred to as nn.}
    \label{fig:edges_dist}
\end{figure}

To further investigate the behavior of coordinated accounts we continue the analysis by performing the classification of edges into three different categories, as explained in Section~\ref{subsec: retweet cascade graph}. Figure~\ref{fig:edges_dist} displays the distribution of the three types of edges: coordinated to coordinated (cc); coordinated to non-coordinated (cn); and non-coordinated to non-coordinated (nn). We mainly observe that across the whole set of cascades the largest proportion of links occurs between non-coordinated accounts while edges among coordinated accounts are quite rare. Nonetheless the two categories are well connected as witnessed by the distribution of the cn edges.

\begin{figure}[h]
    \centering
    \includegraphics[scale = 0.55]{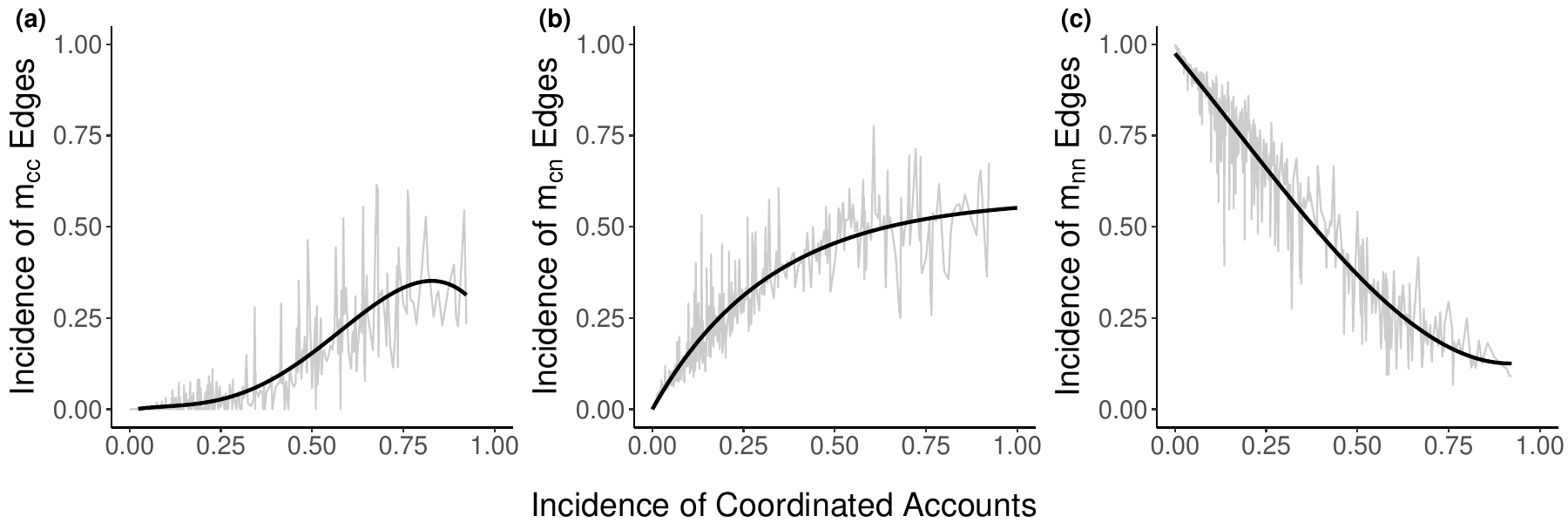}
    \caption{Incidence of different edge types with respect to incidence of coordinated accounts. Panel (a) refers to edges among coordinated accounts; panel (b) refers to edges among coordinated and non-coordinated accounts; panel (c) refers to edges among non-coordinated accounts.}
    \label{fig:loess}
\end{figure}

Moving forward, we also evaluate how the proportion of different edge types evolves with respect to the incidence of coordinated accounts within information cascades. 
Figure~\ref{fig:loess} displays the trends of edges incidence obtained by means of curve fitting.
Panels (a) and (b) report somewhat expected results. The incidence of cc edges grows with the incidence of coordinated accounts (panel (a)), while the opposite trend can be found in the case of nn edges (panel (c)). Interestingly, the curve has a steeper trend in the latter case, indicating that links between non-coordinated accounts tend to decrease rapidly when the percentage of coordinated accounts grows, most likely because of a contextual growth of links between the two categories. Anyway, the most interesting result is that reported in Figure~\ref{fig:loess} (b) where we observe a somewhat linear growth of cn edges up to a saturation point that occurs at $\sim50\%$ (on both axes). 

The presence of a saturation point is confirmed by an appropriate fitting of an exponential model of the form $y = b (1-e^{-ax})$. In such a model, parameter $a$ (typically $a> 1$) determines the steepness of the exponential curve while parameter $b$ is a scaling factor. By performing non-linear least-squares fitting we find that the best fitting values of the parameters are $a =3.100$ and $b=0.578$ with a residual standard error of 0.0732 (on 357 degrees of freedom). The other curves are more conveniently fitted using polynomial functions of the third order in the case of panel~\ref{fig:loess} (a) and of the fourth order in the case of panel~\ref{fig:loess} (c). The selection of polynomials' order was made by means of ANOVA on nested models searching for the lowest order polynomial able to significantly explain the variance of the data, as explained in~\mbox{\citep{james2013introduction}}. The two curves are modeled by the following equations $y = 0.974-1.208x-0.386x^2+0.756x^3$ and $y =-0.002 + 0.168x -1.112x^2 +4.447x^3 -3.291x^4$ for which we find a residual standard error of 0.079 (on 355 degrees of freedom), and 0.068 (on 354 degrees of freedom) respectively.

In practice, the result concerning the saturation point indicates that the incidence of coordinated accounts and, in turn, their capacity to interact with, and to influence, non-coordinated ones, is effective only up to a certain threshold value that we identify at about 50\% of the size of the cascade. Past that point, injecting other coordinated accounts in the network yields no additional benefits, in terms of their capacity to involve non-coordinated accounts in the cascade. In turn, this finding suggests that the targeted use of a limited number of coordinated accounts is a particularly effective manipulation tactic, while the employment of a large number of coordinated accounts appears to be inefficient and ineffective, past the saturation point. Therefore, these results suggest that particular attention should be devoted to the detection of even small groups of coordinated accounts, which calls for the development of accurate and sensitive detectors to support the work of online moderators.

\section{Research Implications}
\label{sec:research implic}

The results presented in this study have important implications in the characterization and contrast of coordinated inauthentic and harmful online behavior. In particular, the peculiar position and activity of coordinated accounts acting within an information cascade, which we highlighted in our work, could be used as further signals for differentiating between authentic and inauthentic online coordination -- an issue that is still largely unsolved~\mbox{\citep{vargas2020detection,nizzoli2021coordinated}}.
In practice, discovering further features that characterize coordinated accounts, beyond their exceptional similarity, could be part of the inputs used by decision support systems that aim at detecting online coordination and manipulated discussions. To this end, our results allowed to identify both node-level and cascade-level features. The identification of such features is an advancement with respect to the current scientific literature on this issue that, to our knowledge, is mainly focused on the development of coordination-detection algorithms~\mbox{\citep{magelinski2021synchronized}}. Our results are thus orthogonal to the existing literature and could support current efforts for developing effective decision support systems for detecting and protecting against coordinated online information manipulations~\mbox{\citep{lozano2020veracity}}.

Another implication of our research involves the presence of a saturation process in the ability of coordinated accounts to involve non-coordinated ones. This indicates that even an imperfect detection of the number of coordinated accounts provides a rather precise estimate of the number of non-coordinated ones. Furthermore, the saturation process signals the ineffectiveness of coordinated actions after a certain threshold value that starts at nearly a half of the cascade size. In turn, a better understanding of the tactics adopted for information manipulation, and their effectiveness, can contribute to prioritize and direct moderation interventions against the coordinated accounts and the manipulated discussions that have the potential of influencing a large number of unaware (non-coordinated) users.

Our work also opens additional research questions. In particular, our results highlighted the challenge of distinguishing between successful coordinated activities, which manage to infect large numbers of non-coordinated users, and non-coordinated activities. The latter, in fact, would appear similar to the former with respect to the number and incidence of coordinated accounts in the information cascade. This calls for additional research and experimentation.


\section{Conclusions}
\label{sec:concl}

In this work we have investigated the role of coordinated accounts in information spreading on Twitter considering the case of the 2019 UK political elections. In particular, we have considered a set of about 50,000 information cascades reconstructed using information deriving from the retweeting activity and from the follower/following network. Overall, we found a small but significant difference in the behavior of coordinated and non-coordinated accounts in terms of node-level measures, namely: positioning level, action delay and descendants count. Coordinated accounts tend to occupy higher positions in the information cascade, share information more rapidly and infect a higher number of users. Furthermore, their incidence in information cascades, especially of small size, seems to derive from strategic actions as confirmed by the departure of the real data from ad-hoc statistical null models. The presence of a strategy is consistent with the definition of coordinated accounts. We tested the ability of coordinated accounts in involving other users by means of novel indices (called infectivity ratios) and link classification in information cascades. In summary, the results of our analyses show that the ability of coordinated accounts to involve non-coordinated ones follows a saturation-like process characterized by a threshold after which injecting more coordinated accounts within a cascade yields negligible effects.




\bibliographystyle{elsarticle-num}
\bibliography{biblio}

\begin{thebibliography}{10}
\expandafter\ifx\csname url\endcsname\relax
  \def\url#1{\texttt{#1}}\fi
\expandafter\ifx\csname urlprefix\endcsname\relax\def\urlprefix{URL }\fi
\expandafter\ifx\csname href\endcsname\relax
  \def\href#1#2{#2} \def\path#1{#1}\fi

\bibitem{bakshy2012role}
E.~Bakshy, I.~Rosenn, C.~Marlow, L.~Adamic, The role of social networks in information diffusion, in: The 21st International Conference on World Wide Web (WWW'12), 2012, pp. 519--528.

\bibitem{flaxman2016filter}
S.~Flaxman, S.~Goel, J.~M. Rao, Filter bubbles, echo chambers, and online news consumption, Public Opinion Quarterly 80~(S1) (2016) 298--320.

\bibitem{schmidt2017anatomy}
A.~L. Schmidt, F.~Zollo, M.~Del~Vicario, A.~Bessi, A.~Scala, G.~Caldarelli, H.~E. Stanley, W.~Quattrociocchi, Anatomy of news consumption on {Facebook}, Proceedings of the National Academy of Sciences 114~(12) (2017) 3035--3039.

\bibitem{hagen2020rise}
L.~Hagen, S.~Neely, T.~E. Keller, R.~Scharf, F.~E. Vasquez, Rise of the machines? examining the influence of social bots on a political discussion network, Social Science Computer Review (2020) 0894439320908190.

\bibitem{trujillo2022make}
A.~Trujillo, S.~Cresci, {Make Reddit Great Again}: Assessing community effects of moderation interventions on {r/The\_Donald}, arXiv preprint arXiv:2201.06455 (2022).

\bibitem{zinovyeva2020antisocial}
E.~Zinovyeva, W.~K. H{\"a}rdle, S.~Lessmann, Antisocial online behavior detection using deep learning, Decision Support Systems 138 (2020) 113362.

\bibitem{santos2021link}
F.~P. Santos, Y.~Lelkes, S.~A. Levin, Link recommendation algorithms and dynamics of polarization in online social networks, Proceedings of the National Academy of Sciences 118~(50) (2021).

\bibitem{hristakieva2022spread}
K.~Hristakieva, S.~Cresci, G.~D.~S. Martino, M.~Conti, P.~Nakov, The spread of propaganda by coordinated communities on social media, in: The 14th International ACM Web Science Conference (WebSci'22), ACM, 2022.

\bibitem{cinelli2020limited}
M.~Cinelli, S.~Cresci, A.~Galeazzi, W.~Quattrociocchi, M.~Tesconi, {The limited reach of fake news on Twitter during 2019 European elections}, PLoS ONE 15~(6) (2020) e0234689.

\bibitem{yuan2021improving}
H.~Yuan, J.~Zheng, Q.~Ye, Y.~Qian, Y.~Zhang, Improving fake news detection with domain-adversarial and graph-attention neural network, Decision Support Systems (2021) 113633.

\bibitem{del2016spreading}
M.~Del~Vicario, A.~Bessi, F.~Zollo, F.~Petroni, A.~Scala, G.~Caldarelli, H.~E. Stanley, W.~Quattrociocchi, The spreading of misinformation online, Proceedings of the National Academy of Sciences 113~(3) (2016) 554--559.

\bibitem{lazer2018science}
D.~M. Lazer, M.~A. Baum, Y.~Benkler, A.~J. Berinsky, K.~M. Greenhill, F.~Menczer, M.~J. Metzger, B.~Nyhan, G.~Pennycook, D.~Rothschild, M.~Schudson, S.~Sloman, C.~Sunstein, E.~A. Thorson, D.~J. Watts, J.~L. Zittrain, The science of fake news, Science 359~(6380) (2018) 1094--1096.

\bibitem{zhang2020overview}
X.~Zhang, A.~A. Ghorbani, An overview of online fake news: Characterization, detection, and discussion, Information Processing \& Management 57~(2) (2020) 102025.

\bibitem{shao2018spread}
C.~Shao, G.~L. Ciampaglia, O.~Varol, K.-C. Yang, A.~Flammini, F.~Menczer, The spread of low-credibility content by social bots, Nature communications 9~(1) (2018) 1--9.

\bibitem{mendoza2020bots}
M.~Mendoza, M.~Tesconi, S.~Cresci, {Bots in social and interaction networks: Detection and impact estimation}, ACM Transactions on Information Systems 39~(1) (2020) 1--32.

\bibitem{boneh2019relevant}
D.~Boneh, A.~J. Grotto, P.~McDaniel, N.~Papernot, How relevant is the turing test in the age of sophisbots?, IEEE Security \& Privacy 17~(6) (2019) 64--71.

\bibitem{starbird2019disinformation}
K.~Starbird, Disinformation's spread: bots, trolls and all of us, Nature 571~(7766) (2019) 449--450.

\bibitem{weedon2017information}
J.~Weedon, W.~Nuland, A.~Stamos, Information operations and {Facebook}, Tech. rep., Facebook (2017).

\bibitem{nizzoli2021coordinated}
L.~Nizzoli, S.~Tardelli, M.~Avvenuti, S.~Cresci, M.~Tesconi, {Coordinated behavior on social media in 2019 UK General Election}, in: The 15th International AAAI Conference on Web and Social Media (ICWSM'21), AAAI, 2021.

\bibitem{pacheco2020uncovering}
D.~Pacheco, P.-M. Hui, C.~Torres-Lugo, B.~T. Truong, A.~Flammini, F.~Menczer, Uncovering coordinated networks on social media, in: The 15th International AAAI Conference on Web and Social Media (ICWSM'21), AAAI, 2021.

\bibitem{firdaus2018retweet}
S.~N. Firdaus, C.~Ding, A.~Sadeghian, Retweet: A popular information diffusion mechanism -- a survey paper, Online Social Networks and Media 6 (2018) 26--40.

\bibitem{ren2019generalized}
X.-L. Ren, N.~Gleinig, D.~Helbing, N.~Antulov-Fantulin, Generalized network dismantling, Proceedings of the National Academy of Sciences 116~(14) (2019) 6554--6559.

\bibitem{vargas2020detection}
L.~Vargas, P.~Emami, P.~Traynor, On the detection of disinformation campaign activity with network analysis, in: Proceedings of the 2020 ACM SIGSAC Conference on Cloud Computing Security Workshop (SIGSAC'20), 2020, pp. 133--146.

\bibitem{lozano2020veracity}
M.~G. Lozano, J.~Brynielsson, U.~Franke, M.~Rosell, E.~Tj{\"o}rnhammar, S.~Varga, V.~Vlassov, Veracity assessment of online data, Decision Support Systems 129 (2020) 113132.

\bibitem{magelinski2021synchronized}
T.~Magelinski, L.~H.~X. Ng, K.~M. Carley, A synchronized action framework for responsible detection of coordination on social media, arXiv preprint arXiv:2105.07454 (2021).

\bibitem{tardelli2021detecting}
S.~Tardelli, M.~Avvenuti, M.~Tesconi, S.~Cresci, Detecting inorganic financial campaigns on {Twitter}, Information Systems (2021) 101769.

\bibitem{alassad2020combining}
M.~Alassad, B.~Spann, N.~Agarwal, Combining advanced computational social science and graph theoretic techniques to reveal adversarial information operations, Information Processing \& Management 58~(1) (2021) 102385.

\bibitem{weber2020s}
D.~Weber, F.~Neumann, Who's in the gang? {Revealing} coordinating communities in social media, in: The 2020 IEEE/ACM International Conference on Advances in Social Networks Analysis and Mining (ASONAM'20), IEEE, 2020, pp. 89--93.

\bibitem{ng2021coordinating}
L.~H.~X. Ng, I.~Cruickshank, K.~M. Carley, Coordinating narratives and the capitol riots on parler, arXiv preprint arXiv:2109.00945 (2021).

\bibitem{schoch2022coordination}
D.~Schoch, F.~B. Keller, S.~Stier, J.~Yang, Coordination patterns reveal online political astroturfing across the world, Scientific reports 12~(1) (2022) 1--10.

\bibitem{giglietto2020takes}
F.~Giglietto, N.~Righetti, L.~Rossi, G.~Marino, It takes a village to manipulate the media: {Coordinated} link sharing behavior during 2018 and 2019 {Italian} elections, Information, Communication \& Society 23~(6) (2020) 867--891.

\bibitem{zhang2021vigdet}
Y.~Zhang, K.~Sharma, Y.~Liu, Vigdet: Knowledge informed neural temporal point process for coordination detection on social media, Advances in Neural Information Processing Systems (NeurIPS'21) 34 (2021).

\bibitem{sharma2021identifying}
K.~Sharma, Y.~Zhang, E.~Ferrara, Y.~Liu, Identifying coordinated accounts on social media through hidden influence and group behaviours, in: The 27th ACM SIGKDD Conference on Knowledge Discovery and Data Mining (KDD'21), ACM, 2021, pp. 1441--1451.

\bibitem{weber2021amplifying}
D.~Weber, F.~Neumann, Amplifying influence through coordinated behaviour in social networks, Social Network Analysis and Mining 11~(1) (2021) 1--42.

\bibitem{ferrara2016rise}
E.~Ferrara, O.~Varol, C.~Davis, F.~Menczer, A.~Flammini, The rise of social bots, Communications of the ACM 59~(7) (2016) 96--104.

\bibitem{fornacciari2018holistic}
P.~Fornacciari, M.~Mordonini, A.~Poggi, L.~Sani, M.~Tomaiuolo, A holistic system for troll detection on {Twitter}, Computers in Human Behavior 89 (2018) 258--268.

\bibitem{kang2021can}
S.~Kang, X.~J. Liu, Y.~Kim, V.~Yoon, Can bots help create knowledge? the effects of bot intervention in open collaboration, Decision Support Systems 148 (2021) 113601.

\bibitem{stella2018bots}
M.~Stella, E.~Ferrara, M.~De~Domenico, Bots increase exposure to negative and inflammatory content in online social systems, Proceedings of the National Academy of Sciences 115~(49) (2018) 12435--12440.

\bibitem{vosoughi2018spread}
S.~Vosoughi, D.~Roy, S.~Aral, The spread of true and false news online, Science 359~(6380) (2018) 1146--1151.

\bibitem{woolley2016automating}
S.~C. Woolley, Automating power: Social bot interference in global politics, First Monday (2016).

\bibitem{mirtaheri2021identifying}
M.~Mirtaheri, S.~Abu-El-Haija, F.~Morstatter, G.~Ver~Steeg, A.~Galstyan, Identifying and analyzing cryptocurrency manipulations in social media, IEEE Transactions on Computational Social Systems 8~(3) (2021) 607--617.

\bibitem{yuan2019examining}
X.~Yuan, R.~J. Schuchard, A.~T. Crooks, Examining emergent communities and social bots within the polarized online vaccination debate in {Twitter}, Social Media+ Society 5~(3) (2019) 2056305119865465.

\bibitem{mazza2019rtbust}
M.~Mazza, S.~Cresci, M.~Avvenuti, W.~Quattrociocchi, M.~Tesconi, {RTbust: Exploiting temporal patterns for botnet detection on Twitter}, in: The 11th International ACM Web Science Conference (WebSci'19), ACM, 2019, pp. 183--192.

\bibitem{cresci2018reaction}
S.~Cresci, M.~Petrocchi, A.~Spognardi, S.~Tognazzi, From reaction to proaction: Unexplored ways to the detection of evolving spambots, in: Companion Proceedings of the The Web Conference 2018 (WWW'18), 2018, pp. 1469--1470.

\bibitem{rauchfleisch2020false}
A.~Rauchfleisch, J.~Kaiser, The false positive problem of automatic bot detection in social science research, PLoS ONE 15~(10) (2020) 1--20.

\bibitem{vosoughi2017rumor}
S.~Vosoughi, M.~N. Mohsenvand, D.~Roy, Rumor gauge: {Predicting the veracity of rumors on Twitter}, ACM transactions on knowledge discovery from data (TKDD) 11~(4) (2017) 1--36.

\bibitem{DBLP:conf/www/YangHZSG12}
C.~Yang, R.~C. Harkreader, J.~Zhang, S.~Shin, G.~Gu, Analyzing spammers' social networks for fun and profit: a case study of cyber criminal ecosystem on {Twitter}, in: Proceedings of the 21st World Wide Web Conference (WWW'12), ACM, 2012, pp. 71--80.

\bibitem{DBLP:journals/corr/zaman}
T.~Zaman, E.~B. Fox, E.~T. Bradlow, A bayesian approach for predicting the popularity of tweets, Annals of Applied Statistics 8~(3) (2014) 1583--1611.

\bibitem{DBLP:conf/cikm/CaoSCOC17}
Q.~Cao, H.~Shen, K.~Cen, W.~Ouyang, X.~Cheng, Deephawkes: Bridging the gap between prediction and understanding of information cascades, in: Proceedings of the 2017 Conference on Information and Knowledge Management (CIKM'17), {ACM}, 2017, pp. 1149--1158.

\bibitem{DBLP:conf/sitis/CazabetPTT13}
R.~Cazabet, N.~Pervin, F.~Toriumi, H.~Takeda, Information diffusion on {Twitter}: Everyone has its chance, but all chances are not equal, in: Proceedings of the 9th International Conference on Signal-Image Technology {\&} Internet-Based Systems (SITIS'13), IEEE, 2013, pp. 483--490.

\bibitem{DBLP:conf/www/TaxidouF14}
I.~Taxidou, P.~M. Fischer, Online analysis of information diffusion in {Twitter}, in: The 23rd International World Wide Web Conference (WWW'14), {ACM}, 2014, pp. 1313--1318.

\bibitem{yang2010predicting}
J.~Yang, S.~Counts, Predicting the speed, scale, and range of information diffusion in {Twitter}, in: Proceedings of the International AAAI Conference on Web and Social Media (ICWSM'10), Vol.~4, 2010.

\bibitem{DBLP:journals/tkde/WuCZCLM20}
B.~Wu, W.-H. Cheng, Y.~Zhang, J.~Cao, J.~Li, T.~Mei, Unlocking author power: On the exploitation of auxiliary author-retweeter relations for predicting key retweeters, IEEE Transactions on Knowledge and Data Engineering 32~(3) (2018) 547--559.

\bibitem{DBLP:conf/worldcist/RodriguesCIPS16}
T.~Rodrigues, T.~Cunha, D.~Ienco, P.~Poncelet, C.~Soares, {RetweetPatterns: Detection of spatio-temporal patterns of retweets}, in: New Advances in Information Systems and Technologies, Springer, 2016, pp. 879--888.

\bibitem{zola2020interaction}
P.~Zola, G.~Cola, M.~Mazza, M.~Tesconi, Interaction strength analysis to model retweet cascade graphs, Applied Sciences 10~(23) (2020) 8394.

\bibitem{de2015towards}
T.~De~Nies, I.~Taxidou, A.~Dimou, R.~Verborgh, P.~M. Fischer, E.~Mannens, R.~Van~de Walle, Towards multi-level provenance reconstruction of information diffusion on social media, in: Proceedings of the 24th ACM International Conference on Information and Knowledge Management (CIKM'15), 2015, pp. 1823--1826.

\bibitem{park2007distribution}
J.~Park, A.-L. Barab{\'a}si, Distribution of node characteristics in complex networks, Proceedings of the National Academy of Sciences 104~(46) (2007) 17916--17920.

\bibitem{cinelli2020network}
M.~Cinelli, L.~Peel, A.~Iovanella, J.-C. Delvenne, Network constraints on the mixing patterns of binary node metadata, Physical Review E 102~(6) (2020) 062310.

\bibitem{zang2017quantifying}
C.~Zang, P.~Cui, C.~Song, C.~Faloutsos, W.~Zhu, Quantifying structural patterns of information cascades, in: Proceedings of the 26th International Conference on World Wide Web Companion (WWW'17 Companion), 2017, pp. 867--868.

\bibitem{giatsoglou2015retweeting}
M.~Giatsoglou, D.~Chatzakou, N.~Shah, C.~Faloutsos, A.~Vakali, Retweeting activity on {Twitter}: Signs of deception, in: Pacific-Asia Conference on Knowledge Discovery and Data Mining (PAKKD'15), Springer, 2015, pp. 122--134.

\bibitem{vo2017revealing}
N.~Vo, K.~Lee, C.~Cao, T.~Tran, H.~Choi, Revealing and detecting malicious retweeter groups, in: The 2017 IEEE/ACM International Conference on Advances in Social Networks Analysis and Mining (ASONAM'17), IEEE, 2017, pp. 363--368.

\bibitem{gupta2019malreg}
S.~Gupta, P.~Kumaraguru, T.~Chakraborty, Malreg: Detecting and analyzing malicious retweeter groups, in: Proceedings of the ACM India Joint International Conference on Data Science and Management of Data (CoDS-COMAD'19), 2019, pp. 61--69.

\bibitem{james2013introduction}
G.~James, D.~Witten, T.~Hastie, R.~Tibshirani, An introduction to statistical learning, Vol. 112, Springer, 2013.

\end{thebibliography}

\end{document}